\documentclass[a4paper]{aa}

\usepackage{graphicx}

\def \rsun {\ifmmode$R$_{\odot}\else R$_{\odot}$\fi}

\def \hcm {\hbox {\ifmmode $ atoms cm$^{-2}\else atoms cm$^{-2}$\fi}}
\def \src {Her\,X-1}
\def\approxgt{\mathrel{\hbox{\rlap{\lower.55ex \hbox {$\sim$}}
        \kern-.3em \raise.4ex \hbox{$>$}}}}
\def\approxlt{\mathrel{\hbox{\rlap{\lower.55ex \hbox {$\sim$}}
        \kern-.3em \raise.4ex \hbox{$<$}}}}

\newcommand{\SAX}{BeppoSAX}

\newcommand {\chisq} {$\chi ^{2}$}

\begin{document}

\title{A BeppoSAX observation of Her X-1 during the first main-on after
an anomalous low-state: evidence for rapid spin-down}

\author{T.~Oosterbroek\inst{1}, A. N.~Parmar\inst{1},
M. Orlandini\inst{2}, A. Segreto\inst{3}, A. Santangelo\inst{3}, and S. Del Sordo\inst{3}}

\offprints{T. Oosterbroek (toosterb@astro.estec.esa.nl)}
\institute{Astrophysics Division, Space Science Department of ESA, 
ESTEC, P.O. Box 299, 2200 AG Noordwijk, The Netherlands
\and Istituto TESRE, CNR, via Gobetti 101, I-40129 Bologna, Italy
\and IFCAI, CNR, via La Malfa 153, I-90146 Palermo, Italy}


\date{Received 17 April 2001 / Accepted 11 June 2001}

\authorrunning{T. Oosterbroek et al.}
\titlerunning{\SAX\ observation of Her X-1 shortly after a long anomalous
low state}

\abstract{Results of a \SAX\ observation of \src\ in 2000 October during the
first main-on state after the longest recorded anomalous low-state 
are presented.
The 0.1--30~keV spectrum, light curve and pulse profile are 
all consistent with those measured during previous main on-states, 
indicating that \src\ has resumed its regular 35 day
cycle with similar on-state properties as before. 
However, from a comparison of the measured pulse period with that obtained
close to the start of the anomalous low-state, 
it is evident that \src\ continued
to spin-down strongly during the anomalous low-state such that
the pulse period has returned to a similar value as $\sim$15~years ago.
Additionally, the occurrence time of the main-on states after the end of the
anomalous low-state indicate that a change in the
length, or phasing, of the 35-day cycle occurred during the anomalous 
low-state.
\keywords{accretion, accretion
disks -- X-rays: binaries -- individual: \src} \\ } \maketitle

\section{Introduction}

\src\ is an eclipsing binary X-ray pulsar with a pulse period of
1.24s and an orbital period of 1.7~days (Tananbaum et al.\ \cite{t:72};
Giacconi et al. \cite{g:73}). 
Normally, the source exhibits a 35~day X-ray intensity cycle comprising a~
$\sim$10 day duration main on-state and a fainter $\sim$5 day duration 
secondary on-state approximately half a cycle later. At other phases of
the 35~day cycle, \src\ is still visible at a low level (Jones \& Forman
\cite{j:76}). This modulation has been ascribed to a tilted precessing
accretion disk that periodically obscures the line
of sight to the neutron star (Gerend \& Boynton \cite{g:76}).  
In addition, a regular pattern of X-ray 
intensity dips is usually observed at certain orbital phases. These may be
caused by obscuration from periodically released matter from the companion
star (Crosa \& Boynton \cite{c:80}). 

The 35-day cycle has been evident in RXTE All-Sky Monitor (ASM)
1.5--12~keV data with the main-on state being clearly detected every
35 days for $>$3~years (e.g., Scott \& Leahy \cite{s:99}).  An
exception to this regularity occurred when the on-state expected
around 1999 March~23 was not detected (Levine \& Corbet \cite{l:99}).
Observations with \SAX\ showed only low level activity during the
expected time of the next main-on (Parmar et al.\
\cite{p:99}). Similar exceptions have been detected twice before.  In
1983 June to August EXOSAT failed to detect an X-ray on-state from
\src\ and instead a faint source, with a strength comparable to that
of the low-state emission was observed over a wide range of 35-day
phases (Parmar et al. \cite{p:85}). Optical observations during this
interval, and in 1999 April, detected the effects of strong X-ray
heating on the companion star (HZ~Her) indicating that it was still
being irradiated by a strong X-ray source (Delgado et al. \cite{d:83};
Margon et al.  \cite{m:99}).  The 1983 anomalous low-state lasted
$<$0.8~year with the 35-day cycle returning by 1984 March.  Similarly,
in 1993 August ASCA failed to observe the expected on-state, again
detecting instead a faint X-ray source (Vrtilek et al. \cite{v:94};
Mihara \& Soong \cite{m:94}).  The latest anomalous low-state is the
longest ever observed, lasting 1.5~years.  During the beginning of
this anomalous low-state a rapid spin-down of the pulse period was
observed (Parmar et al.\ \cite{p:99}; Coburn et al.\ \cite{c:00}),
which was interpreted as resulting from a torque-reversal, possibly
caused by a reversal of the rotation of the inner disk (see e.g., van
Kerkwijk et al.\ \cite{vk:98}).

\begin{figure*}
\centerline{\includegraphics[width=8cm,angle=-90]{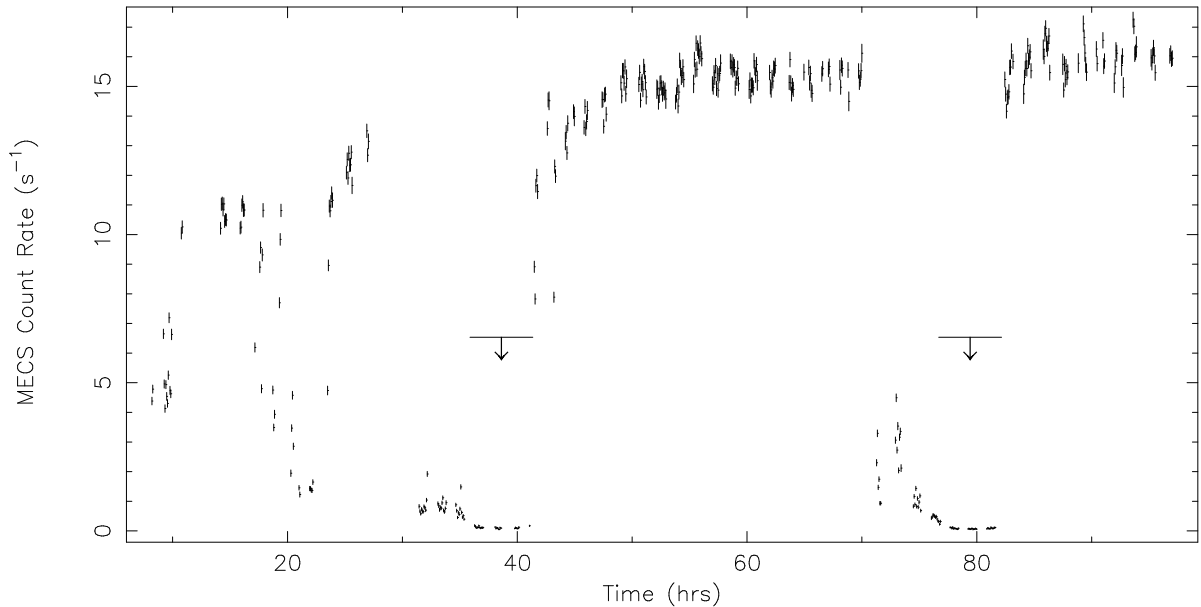}}
\caption[]{The MECS 2--10 keV light curve of \src\ with a binning of 256~s.
The time is hours of 2000 October 9. The eclipse intervals are
indicated by arrows and horizontal lines} 
\label{fig:lc}
\end{figure*}

Here we report on a \SAX\ observation during the first main-on
state (as derived from the RXTE-ASM light curve) after the 1999 March to
2000 September anomalous low-state
had ended.
We discuss the source properties and show that the 0.1--30~keV spectrum,
light curves and pulse profiles are 
indistinguishable from those during previous
high-states, but that the pulse period probably continued its strong spin-down
behavior during the anomalous low-state itself.
Additionally, we examine the 35 day turn-on times after and before the
anomalous low-state and find that the 35 day period, or phasing, changed during the
anomalous low-state. 
\section{Observations}

Results from the Low-Energy Concentrator Spectrometer (LECS;
0.1--10~keV; Parmar et al. \cite{p:97}), the Medium-Energy Concentrator
Spectrometer (MECS; 1.8--10~keV; Boella et al. \cite{b:97}),
and the Phoswich Detection System (PDS; 15--300~keV; 
Frontera et al. \cite{f:97}) on-board \SAX\ are presented.
The MECS consists of two grazing incidence
telescopes with imaging gas scintillation proportional counters in
their focal planes. The LECS uses an identical concentrator system as
the MECS, but utilizes an ultra-thin entrance window and
a driftless configuration to extend the low-energy response to
0.1~keV. The non-imaging
PDS consists of four independent units arranged in pairs each having a
separate collimator. Each collimator was alternatively
rocked on- and 210\arcmin\ off-source every 96~s during 
the observation. The HPGSPC was not operated during this observation.

The observation was performed during the rising edge of the main-on
between 2000 October 9 08:12 and October 13
01:08 (UTC). Examination of the RXTE-ASM light curve shows that the
start of the on-state occurred a few hours earlier. This is also seen
in the MECS light curve (Fig.\ \ref{fig:lc}) where the count rate at
the start is already increasing and higher than the low-state value of
$\sim$1~s$^{-1}$ (see Oosterbroek et al.~\cite{o:00}).  The standard
LECS and MECS extraction radii of 8$'$ and 4$'$ were used.  Standard
data selection and background subtraction procedures were applied
using the using the SAXDAS 2.0.0 data analysis package.  The resulting
exposures in the LECS, MECS, and PDS instruments are 28.5~ks,
106.0~ks, and 48.1~ks, respectively.

\section{Analysis}

\begin{figure}
\centerline{\includegraphics[height=7cm,angle=0]{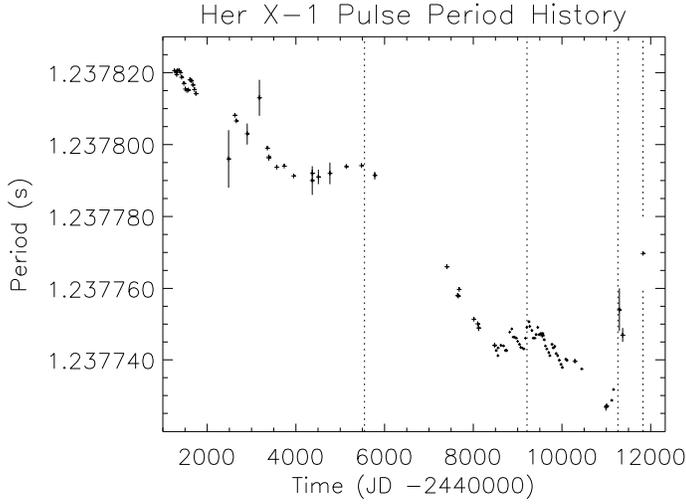}}
\caption[]{The Her X-1 pulse period history. The two leftmost dashed
vertical lines mark the previous anomalous low-states, while the two
rightmost dashed vertical lines indicate the beginning and end of
the latest anomalous low-state}

\label{fig:pulse_history}
\end{figure}

The pulse period of \src\ was determined.
First, the arrival times of the photons were corrected
to the solar system barycenter. Then, the arrival times were
additionally corrected to the \src\ center of mass using the ephemeris
of Deeter et al.\ (\cite{d:91}). 
The period was obtained with an epoch-folding
technique using the MECS data, while the (1$\sigma$) uncertainties
were determined by fitting the arrival
times of 17 averaged profiles. We find a period of 1.2377697(3)~s.
This pulse period is plotted in Fig.\ \ref{fig:pulse_history}
together with a compilation of previously obtained pulse periods (see
Parmar et al.\ \cite{p:99} and  Coburn et al.~(\cite{c:00}) for references).

\begin{figure}
\centerline{\includegraphics[height=7cm,angle=-90]{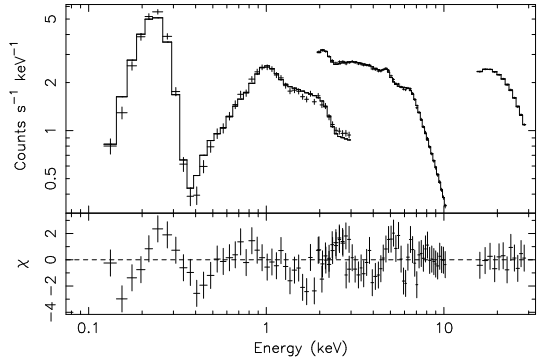}}
\caption[]{The LECS, MECS and PDS count spectra obtained during the
third orbit covering the 0.1--30
keV energy range. The structured low-energy
residuals are caused by the unusually low LECS detector 
temperature which is not properly modeled
by the response generating software}
\label{fig:spectrum}
\end{figure}

\begin{figure}
\centerline{\includegraphics[width=6cm,angle=-90]{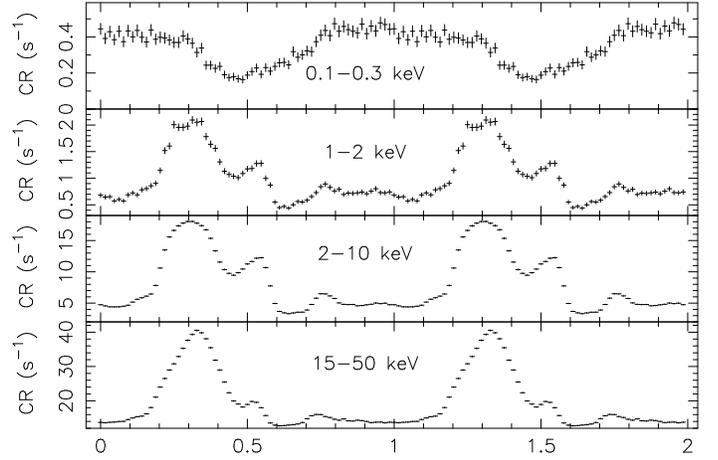}}
\caption[]{The Her X-1 pulse profile in 4 energy bands (indicated in
the panels). Pulse phase is arbitrary but is the same for each panel. 
From top to bottom the data are obtained with the LECS, LECS, MECS
and PDS instruments. The PDS data have not been background subtracted}
\label{fig:profile}
\end{figure}

\begin{figure}
\centerline{\includegraphics[width=8cm]{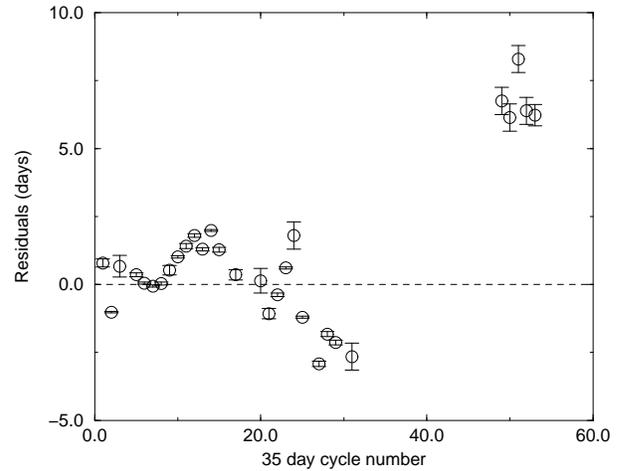}}
\caption[]{The residuals (O-C) of the mean times of the 35 day on-state with
respect to the best-fit linear ephemeris obtained from on-states before the
start of the anomalous low-state}
\label{fig:35day}
\end{figure}

The LECS and MECS spectra were rebinned to have $>$20 counts in each
bin.  Additionally, a rebinning was performed such that the full width
half-maximum energy resolution is oversampled by at most a factor 3 to
ensure the applicability of the \chisq\ statistic.  A 1\% systematic
uncertainty was added to the rebinned data.  Spectra were obtained
during the whole of the second and third orbits (excluding dips and
eclipses) and during 5 small intervals of the first orbit (since rapid
variations are present). The LECS, MECS and PDS spectra were fit with
the model described in Oosterbroek et al. (\cite{o:00}), which
consists of a soft blackbody and a hard power-law, together with two
Gaussian emission features near 6.5 and 1.0 keV (describing two iron
lines). The spectra obtained during the second and third orbits (see
Fig.\ \ref{fig:spectrum}) are in all respects similar to the spectra
obtained during previous main-on states, while the spectrum obtained
during the first orbit can be adequately described with the standard
model modified by partial covering. The spectrum (Fig.\
\ref{fig:spectrum}) was fit using the model in Dal Fiume et al.\
(\cite{df:98}), without the addition of the cyclotron line (since we
limit our spectrum to 30 keV). This model consists of a broken
powerlaw (with a break-energy E$_{\rm break}$, and two power-law
indices, $\alpha_{\rm 1}$ and $\alpha_{\rm 2}$) modified by a
high-energy cutoff (characterized by E$_{\rm cutoff}$, and , E$_{\rm
fold}$) two gaussians, and a (soft) blackbody. The main parameters of
the continuum are (with the Dal Fiume \cite{df:98} model 1 values in
parentheses): $\alpha_{\rm 1}$ $0.89\pm0.01 (0.884\pm0.003)$, E$_{\rm
break}$ (keV) $17.1\pm^{0.3}_{0.8} (17.74\pm^{0.26}_{0.30})$,
$\alpha_{\rm 2}$ $1.86\pm0.4 (1.83\pm^{0.05}_{0.07})$, E$_{\rm cutoff}
$(keV) $ 23.5\pm2 (24.2\pm0.2)$, E$_{\rm fold} $(keV) $ 15.5\pm5
(14.8\pm0.4)$

Therefore, the spectral parameters are completely consistent with
those of the normal main-on spectrum, and we therefore conclude that
the 0.1--30~keV spectrum during this main-on state is in no way
remarkable.

In Fig.\ \ref{fig:profile} the pulse profiles obtained
in 4 different energy bands are plotted. These
clearly show the well known transition from a broad, approximately
sinusoidal, profile at low energies (associated with the soft
blackbody component, see e.g.,\ Oosterbroek et al.\ \cite{o:97}) to the
sharper double-peaked profile at higher energies (associated with the
power-law component). They are very similar to e.g., the main-on
observation reported in Oosterbroek et al.\ (\cite{o:97}).

\subsection{35 day turn-on time}

In order to investigate whether the 35 day on-states observed after
the anomalous low-state occurred at times consistent with their ephemeris
before the low-state RXTE-ASM data covering the interval
1996 January 05 to 2001 February 26 were used. 
Only ASM detections with a 
signal to noise
ratio of $>$7.5 were included (this had the side-effect of excluding all 
but one measurement
outside the main-on state - which was
manually deleted).
The mean occurrence time of each the main-on state intensity was
determined to be the mean of the 2--12~keV data points belonging to each
main-on, while the uncertainty was taken as the the square root
of the variance divided by
the number of points (i.e. the error in the mean). We note that this
is only a very rough estimate of the uncertainty, and does not take
into account any systematic uncertainties. However, our results are
insensitive to the choice of uncertainties.
Whilst this is a very rough method, the results agree
surprisingly well with the more detailed analysis (basically a
template fitting method) of Shakura et al.\ (\cite{s:98}). Next
a linear ephemeris was fit to the times of
the main-ons that occurred before the anomalous low-state to give
a mean 35 day cycle time of 34.8~day, consistent with the
model of Staubert et al.~(\cite{s:83}) where it is proposed that
the mean period is 20.5 times the orbital period, ie., 34.85~day
and that the observed on-states can fluctuate around this value by
0 or $\pm$${1 \over 2}$ orbital cycles. 
This simple ephemeris gives a reasonable
description of the observed main-on times, while the extrapolation to
the main-ons observed {\it after} the anomalous low-state by
the RXTE-ASM show
that these occurred $\sim$7~days (or $\sim$0.2 in phase) late 
(or alternatively, $\sim$0.8 in phase early etc). 
This ambiguity is present because the 35 day cycle count is not
available during the anomalous low-state.
Fig.~\ref{fig:35day} shows the residuals
with respect to the best-fit ephemeris assuming the former case. 
A clear difference between the
on-state occurrence times before and after the anomalous low-state 
is evident.

\section{Discussion}

We have observed \src\ during the first main-on state following
the 1999 March to 2000 September anomalous low-state. We find that 
the 0.1--30~keV light
curves, spectra, and pulse profiles are indistinguishable from
those of previous main-on states and we
therefore conclude that we observed Her X-1 during a normal main-on
state.

The rapid spin-down of Her X-1 continued
during at least part of the anomalous
low-state, such that the pulse period is now at a similar
value to $\sim$15~years ago. This implies a spin-{\it down} rate which
is a factor of $\sim$9 times larger than the average spin-{\it up} rate.
Vrtilek et al.\ (\cite{v:01}) noted that the magnitude of deviation
from spin-down appears to be correlated with the length of the
anomalous low state; this observation confirms this.
There is no indication that the mass accretion rate changed
substantially during the anomalous low-state since optical observations
detected the effects of continued strong X-ray heating on the companion star
(Margon et al.\ \cite{m:99}). 
This implies that the spin-down torque originating from
approximately the same amount of mass transfer was $\sim$9 times 
stronger during the
anomalous low-state than during normal intervals. 
Periods of more rapid spin-up than average are seen
occasionally in \src, but the magnitude of the spin-down rate 
observed during the recent anomalous low state is still larger by
a factor of $\sim$2.5 than the most rapid spin-up rate
previously observed (around
day 7300--8000 in Fig.\ \ref{fig:pulse_history}).

This imposes constraints on any theory modeling the accretion
torques in this source. The low-mass
X-ray binary 4U\thinspace1626-67 underwent a
torque-reversal (accompanied by a change in the spectrum, see
Chakrabarty et al.\ \cite{c:97}), but with only a change in the sign
(not magnitude). This appears substantially different then a
torque-reversal combined with a change in magnitude of a factor $\sim$9
observed here.  
It
has been suggested (Li \& Wickramasinghe \cite{lw:98}) that variations
in the spin-down rate can be caused by changes in the structure of the outer
magnetosphere of the neutron star. In our case these changes may be
accompanied by changes in the disk structure which produce the
anomalous low-state. 

The measured mean times of the main-on states that occur after
the anomalous low-state occur at times inconsistent by at least
$\sim$7~days, or 0.2 in 35 day phase, with a simple linear
extrapolation of the previous on-state times.  However, we cannot
exclude that this is actually 0.8 cycles earlier in 35 day phase, or
some other multiple of the cycle.  This difference may be explained by
either a change in the 35 day phase at which the main on-states occur,
or by a change in the underlying 35 day recurrence interval.  It is
probable that seventeen 35 day cycles occurred during the anomalous
low-state. A change of $\sim$1.5\% in the 35 day period could give
rise to the observed difference.  As \"Ogelman~(\cite{o:87}) points
out, the largest ($-$6.5~day) 35 day phase excursion recorded
previously occurred at the time of the 1983 anomalous low-state. This
suggests that the occurrence of the anomalous low-states and mechanism
responsible for the 35 day cycle in \src\ are somehow linked.  Shakura
et al.\ (\cite{s:98}) report a positive correlation between the length
of the 35 day cycle and the mean X-ray flux.  During the anomalous
low-state the effects of strong X-ray heating on HZ~Her were still
evident (Margon et al.~\cite{m:99}). However, there is no reliable
estimate of how much the mean X-ray flux changed compared to intervals
when the 35 day cycle is present. Indeed, since the mechanism
producing this correlation is unknown, it is unclear whether the
amount of X-ray heating experienced by HZ~Her is a reliable indicator
for this effect.
This correlation suggests that during the anomalous low-state whichever 
part of the
\src\ system regulated this mechanism received a significantly different
mean X-ray intensity than during normal intervals.

\acknowledgements 
The \SAX\ satellite is a joint Italian and Dutch programme.
We thank the staffs of the \SAX\ Science Data and
Operations Control Centers for help with these observations. 
We thank the RXTE instrument teams at MIT and NASA/GSFC for 
providing the All-Sky Monitor data. 

{}
\end{document}